\documentclass{article}
\usepackage{spconf,amsmath,graphicx,hyperref}
\usepackage{booktabs}   
\usepackage{multirow}   
\usepackage{float} 
\usepackage{capt-of}

\title{Utilizing Information Theoretic Approach to Study Cochlear Neural Degeneration}

\name{Ahsan Jamal Cheema$^{1,2}$ \qquad Sunil Puria$^{1,2}$ \thanks{Thanks to NIH-NIDCD (Grant \# R01DC07910 ) for funding.}}
\address{$^{1}$ Harvard University, Speech and Hearing Bioscience and Technology Program, Boston MA, USA\\
$^{2}$ Eaton-Peabody Laboratories, Massachusetts Eye and Ear(MEEI), Boston MA, USA}

\begin{document}
%
\maketitle
\begin{abstract}
Hidden hearing loss, or cochlear neural degeneration (CND), disrupts suprathreshold auditory coding without affecting clinical thresholds, making it difficult to diagnose. We present an information-theoretic framework to evaluate speech stimuli that maximally reveal CND by quantifying mutual information (MI) loss between inner hair cell (IHC) receptor potentials and auditory nerve fiber (ANF) responses and acoustic input and ANF responses. Using a phenomenological auditory model, we simulated responses to 50 CVC words under clean, time-compressed, reverberant, and combined conditions across different presentation levels, with systematically varied survival of low-, medium-, and high-spontaneous-rate fibers. MI was computed channel-wise between IHC and ANF responses and integrated across characteristic frequencies. Information loss was defined relative to a normal-hearing baseline. Results demonstrate progressive MI loss with increasing CND, most pronounced for time-compressed speech, while reverberation produced comparatively smaller effects. These findings identify rapid, temporally dense speech as optimal probes for CND, informing the design of objective clinical diagnostics while revealing problems associated with reverberation as a probe.
\end{abstract}
\begin{keywords}
cochlear synaptopathy, mutual information, hearing loss, hearing aids, information theory
\end{keywords}
%
\section{Introduction}
\label{sec:intro}

One of the most common challenges faced by individuals with sensorineural hearing loss (SNHL) is comprehending speech in noisy environments, even when the sounds are clearly audible \cite{Vermiglio2012-fi}. This difficulty is often expressed as, “I hear you, but I don’t understand you.” While SNHL has long been attributed to dysfunction or loss of cochlear hair cells, animal and human temporal bone studies have demonstrated that synaptic connections between inner hair cells (IHCs) and auditory nerve fibers (ANFs) often degenerate earlier; a phenomenon termed cochlear neural degeneration (CND) or cochlear synaptopathy \cite{Kujawa2009-nv, Liberman2016-ph}. This “hidden hearing loss” does not elevate pure‐tone thresholds and thus escapes detection by standard clinical audiometry \cite{Bharadwaj2015-bl}, yet it can disrupt temporal coding and reduce speech‐in‐noise intelligibility \cite{Bharadwaj2014-te, Plack2014-gc}. It has been hypothesized that even individuals with clinically normal hearing thresholds may experience perceptual deficits due to underlying CND \cite{Liberman2016-ph, Plack2014-gc}. However, current clinical tools lack noninvasive and objective methods for detecting CND or quantifying its perceptual consequences. Clinically, CND is most often inferred indirectly through electrophysiological measures such as diminished auditory brainstem response (ABR) Wave 1 amplitudes \cite{Mehraei2016-vg}, envelope‐following responses (EFRs) \cite{Shaheen2015-hz}, electrocochleography (ECochG) SP/AP ratios \cite{Yasar2025-lo}, and altered middle‐ear muscle reflex thresholds \cite{Valero2018-bh}. While these biomarkers show promise, none provide a direct, quantitative index of suprathreshold information transmission across the full complement of ANF populations, and their sensitivity in human subjects remains a matter of active debate.
To reveal CND, several researchers have advocated for “difficult” speech tests—such as time‑compressed sentences, speech in fluctuating maskers, or rapid amplitude‐modulated signals—that place greater demands on temporal and spectral encoding \cite{Bharadwaj2015-bl}. These paradigms aim to tax low‑ and medium‑spontaneous‑rate fibers, which are most vulnerable to synaptopathy. However, “difficulty” has been defined qualitatively, and there exists no standardized, objective metric to quantify the information demands of a speech stimulus. Without such a framework, normal‑hearing listeners may also struggle on these tasks, confounding attempts to isolate deficits attributable to CND and increasing the risk of false positives in screening. Another challenge in the field has been to develop speech probes that involve minimal central auditory processing such as memory and attention, so that measured performance more directly reflects peripheral encoding fidelity rather than higher-order cognitive factors.

\begin{figure}[t!]
    \centering
    \includegraphics[width=\linewidth]{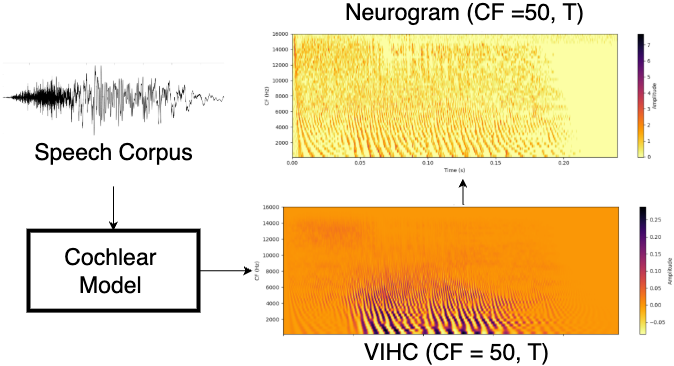} 
    \caption{Schematic of the data generation process. Speech corpus (Section \ref{subsec:speech_corpus}) processed through the cochlear model (Section \ref{subsec: cochlear_model}) to generate inner hair cell receptor potentials and summed spiking activity for all auditory nerve fibers, stored as a 2D matrix and visualized as a time-frequency Neurogram.
    }
    \label{fig:schematic}
\end{figure}

\indent Information theory offers a principled solution by quantifying the mutual information (MI) between stimulus features and neural responses \cite{Cover1991, Nelken2007-ny}. MI has previously been applied to auditory coding in animal models \cite{Louage2004-bp, Gourevitch2006-bq}, model theoretical losses in human auditory periphery using simplified leaky channel assumption \cite{8579632} and to cortical EEG responses in humans \cite{Mikkelsen2017-wk}, but has not yet been used to evaluate peripheral synaptic integrity in a detailed spiking model of auditory periphery. In this study, we introduce an MI‐based framework that computes the degradation in information transmission between IHC receptor potentials and ANF neurograms using a detailed phenomenological model of human auditory periphery \cite{Bruce2018-iv, Zaar2022-sd}. Using the upper limit of the information transmission and by defining information loss as the MI difference between healthy and impaired models, we establish an objective, quantitative measure of stimulus difficulty and CND sensitivity. Our approach is based on a fundamental assumption: if an individual has CND, the speech material used to characterize it should have robust information encoding in normal hearing subjects and the highest information loss in the subjects with CND. Our approach allows us to rank and design of speech materials for hidden hearing loss based on decrease in the theoretical upper limit of information transmission. This approach can be used in future work to design speech materials sensitive to detect CND.
\vspace{-10pt}
\section{Methods}
\label{sec:methods}
\subsection{Speech Corpus}
\label{subsec:speech_corpus}
We used a NU6 List 7 consisting of 50 CVC words as our speech corpus. These words were passed on to the Google Text‑to‑Speech (gTTS) API \cite{GoogleCloudTextToSpeech} and speech material was generated using the most neutral sounding male voice "en-US-Studio-Q", sampled at 44.1 kHz. One hypothesis related to effects of CND is the decrease in information content in the neural signal which becomes more evident only during complex listening tasks \cite{DiNino2022-so}. Therefore in order to simulate these "difficult conditions" \cite{Liberman2016-ph, Grant2022-zs}each word in the corpus was rendered under following four listening conditions of graded difficulty:  
\begin{enumerate}
  \item \emph{Clean speech}: unaltered tokens.  
  \vspace{-6pt}
  \item \emph{40\% Time‑compressed speech}: 60\% time compression applied or original compressed to 40\%.  
  \vspace{-6pt}
  \item \emph{Reverberant speech}: \texttt{reverb\_time}=0.3s and \texttt{decay}=0.3 ( T\textsubscript{60} $\approx$ 0.3s).  
  \vspace{-6pt}
  \item \emph{Combined compression + reverberation}: sequential application of (2) and (3).  
\end{enumerate}
All the speech material were stored as wav files and input to the auditory nerve model \cite{Bruce2018-iv} after normalizing and matching the absolute sound pressure level at 50dB, 65dB, 80dB, 90dB, and 95dB. In total we ended up with 50 words for each presentation level for each speech type. Only results from 90 dB SPL which is suprathershold and where all fibers are recruited are shown in this paper.
\subsection{Phenomenological Model of Auditory Peripheral Processing}
\label{subsec: cochlear_model}
Phenomenological model of cochlea \cite{Bruce2018-iv} was used to simulate the auditory nerve responses to input speech files. The pressure waveform of the speech material was input to the model and the output from the model was in the form of 2D (Characteristic Frequency (CF), Time(T)) matrix of auditory nerve spiking activity. In this study, we utilized fine timing (FT) neurograms to preserve temporal detail and prevent information loss from time averaging hence, throughout the paper, all references to “neurogram” specifically denote FT neurograms consisting of 50 frequency channels and time dimension equal to the length of stimulus. The units of spiking activity are spikes/sec. This model allows simulation of both normal hearing and hearing-impaired conditions. Hearing loss can be modeled by inputting the subject audiogram and model can reduce the gain of cochlear filters based on the input audiogram. The model also allows simulating different levels of CND by reducing the number of auditory nerve fibers per each IHC at a given characteristic frequency (CF). In this study various levels of CND were simulated by reducing the numbers of different ANF types: high-spontaneous (HS) fibers with low thresholds, medium-spontaneous (MS) fibers with intermediate thresholds, and low-spontaneous (LS) fibers with high thresholds. The specific audiometric profiles and synaptopathy configurations used in this study are shown in Tables \ref{tab:audiometric_profile_A} and \ref{tab:CND_profiles}, respectively. Additionally we also record the output of Inner Hair Cell (IHC) receptor potentials from each cochlear filter. A schematic of the neurogram and receptor potential generation pipeline is shown in the Fig. \ref{fig:schematic}.
\vspace{-10pt}
\subsection{Mutual Information Analysis}
\label{subsec:MI_analysis}

An acoustic waveform ($X$), represented as a time series, enters the cochlea where it is decomposed by a bank of frequency-tuned cochlear filters, or characteristic frequencies (CFs). Each filter generates a time series of inner hair cell (IHC) receptor potentials, $V_{\mathrm{IHC}}$, which provide a partially redundant representation of the input due to the spectral overlap of adjacent filters. In this study, we quantify information transmission at two stages. First, we calculate the mutual information (MI) between each IHC receptor-potential time series and the corresponding auditory nerve (AN) neurogram, thereby capturing the fidelity of synaptic transmission from IHCs to ANFs at each CF. This measure is particularly useful because it eliminates the confounding influence of audiometric loss, which primarily reduces cochlear filters and IHC responses. Second, we calculate the MI between the input waveform $X$ and the AN neurogram, which reflects the overall encoding capacity of the periphery, including degradations introduced both at the IHC stage and during subsequent synaptic transmission. Extending these analyses across all CFs provides a channel-wise distribution of information transmission. Because redundancy across overlapping filters is not explicitly removed, these values represent the maximum possible information capacity of the auditory periphery. This upper-bound framework enables us to estimate the maximum potential loss of information that may result from cochlear neural degeneration (CND) or hearing loss. 

Channel-wise mutual information was computed using a histogram-based estimator with $B=1024$ bins (selected via bias–variance trade-off analysis). For a given cochlear channel $f$, the general MI formulation is given by
\begin{equation}
I_f^{(Z \to \mathrm{AN})}
= \sum_{i=1}^{B} \sum_{j=1}^{B}
\hat{p}\!\left(N_f^{(j)},\, Z_f^{(i)}\right)
\log_{2}
\frac{\hat{p}\!\left(N_f^{(j)},\, Z_f^{(i)}\right)}
{\hat{p}\!\left(N_f^{(j)}\right)\,\hat{p}\!\left(Z_f^{(i)}\right)} ,
\label{eq:MI}
\end{equation}
where $N_f$ denotes the AN neurogram in channel $f$, and $Z_f$ is a placeholder variable that can take two forms: $Z_f = V_{\mathrm{IHC},f}$ (IHC receptor potential) for the IHC$\to$AN case, or $Z_f = X_f$ (stimulus representation) for the $X\to$AN case. The estimated distributions $\hat{p}(\cdot)$ represent empirical marginals. The collection of CF-wise MI values yields a vector representation,
\begin{equation}
\mathbf{I}^{(Z \to \mathrm{AN})}
\;=\;
\bigl[\, I_{1}^{(Z \to \mathrm{AN})},\; I_{2}^{(Z \to \mathrm{AN})},\; \ldots,\; I_{N_{\mathrm{CF}}}^{(Z \to \mathrm{AN})} \,\bigr],
\label{eq:Ivec}
\end{equation}
which characterizes the distribution of information across each CF. To summarize these channel-wise values into a single metric per profile, we compute the area under the MI curve (AUC) over log-frequency, weighted by $\log(f)$:
\begin{equation}
\mathrm{AUC}^{(Z \to \mathrm{AN})}
\;=\;
\int_{\log f_{\min}}^{\log f_{\max}}
I^{(Z \to \mathrm{AN})}(f)\,\log(f)\; d\!\bigl(\log f\bigr).
\label{eq:AUC}
\end{equation}
We used 50 CFs therefore, this integral can be approximated as below using the Simpson's rule:
\begin{equation}
\mathrm{AUC}^{(Z \to \mathrm{AN})}
\;\approx\;
\sum_{c=1}^{N_{\mathrm{CF}}}
I_{c}^{(Z \to \mathrm{AN})}\,\log(f_c)\,\Delta\!\bigl(\log f_c\bigr),
\label{eq:AUCdisc}
\end{equation}
where $f_c$ are the characteristic frequencies, $\Delta(\log f_c)$ is the spacing in log-frequency, and $I_c^{(Z \to \mathrm{AN})}$ is the MI at channel $c$.

Finally, to quantify information loss relative to a normal-hearing baseline, we computed the difference in AUC values between each profile $k$ and the normal-hearing (NH) condition:
\begin{equation}
\Delta \mathrm{AUC}^{(k, Z \to \mathrm{AN})}
\;=\;
\mathrm{AUC}^{(k, Z \to \mathrm{AN})}
\;-\;
\mathrm{AUC}^{(\mathrm{NH}, Z \to \mathrm{AN})},
\label{eq:deltaAUC}
\end{equation}
where $k$ indexes the synaptopathy profile (Table~\ref{tab:CND_profiles}).

\begin{figure}[t]
\centering

\begin{minipage}[t]{\columnwidth}\vspace{0pt}%
\footnotesize\setlength{\tabcolsep}{4pt}\renewcommand{\arraystretch}{1.1}
\captionof{table}{Hearing Loss profile used in the study}
\label{tab:audiometric_profile_A}
\resizebox{\columnwidth}{!}{%
\begin{tabular}{lccccccc}
\toprule
Audiometric Profile & 125 Hz & 250 Hz & 500 Hz & 1000 Hz & 2000 Hz & 4000 Hz & 8000 Hz \\
\midrule
Sloping Loss        & 0      & 0      & 10     & 20      & 23      & 45      & 75      \\
\bottomrule
\end{tabular}%
}
\end{minipage}

\vspace{0.75em} 

\begin{minipage}[t]{\columnwidth}\vspace{0pt}%
\footnotesize\setlength{\tabcolsep}{4pt}\renewcommand{\arraystretch}{1.1}
\captionof{table}{CND profiles for the age-related hearing loss in Table~\ref{tab:audiometric_profile_A}. Distribution [LS, MS, HS] = [5, 5, 12] represents fiber counts per inner hair cell in a normal cochlea with no CND.}
\label{tab:CND_profiles}
\resizebox{\columnwidth}{!}{%
\begin{tabular}{llccc}
\toprule
Audiometric Profile & CND Profile                       & Low-SR Fibers & Med-SR Fibers & High-SR Fibers \\
\midrule
\multirow{6}{*}{Sloping Loss} 
 & No CND                          & 5 & 5 & 12 \\
 & 40\% LS MS loss                 & 3 & 3 & 12 \\
 & 80\% LS MS loss                 & 1 & 1 & 12 \\
 & 100\% LS MS loss                & 0 & 0 & 12 \\
 & 100\% LS MS loss, 40\% HS loss  & 0 & 0 & 7  \\
\bottomrule
\end{tabular}%
}
\end{minipage}

\vspace{-1mm} 
\end{figure}

\vspace{-10pt}
\section{Results}
\label{sec:results}
\begin{figure}[t]
    \centering
    \includegraphics[width=1\linewidth]{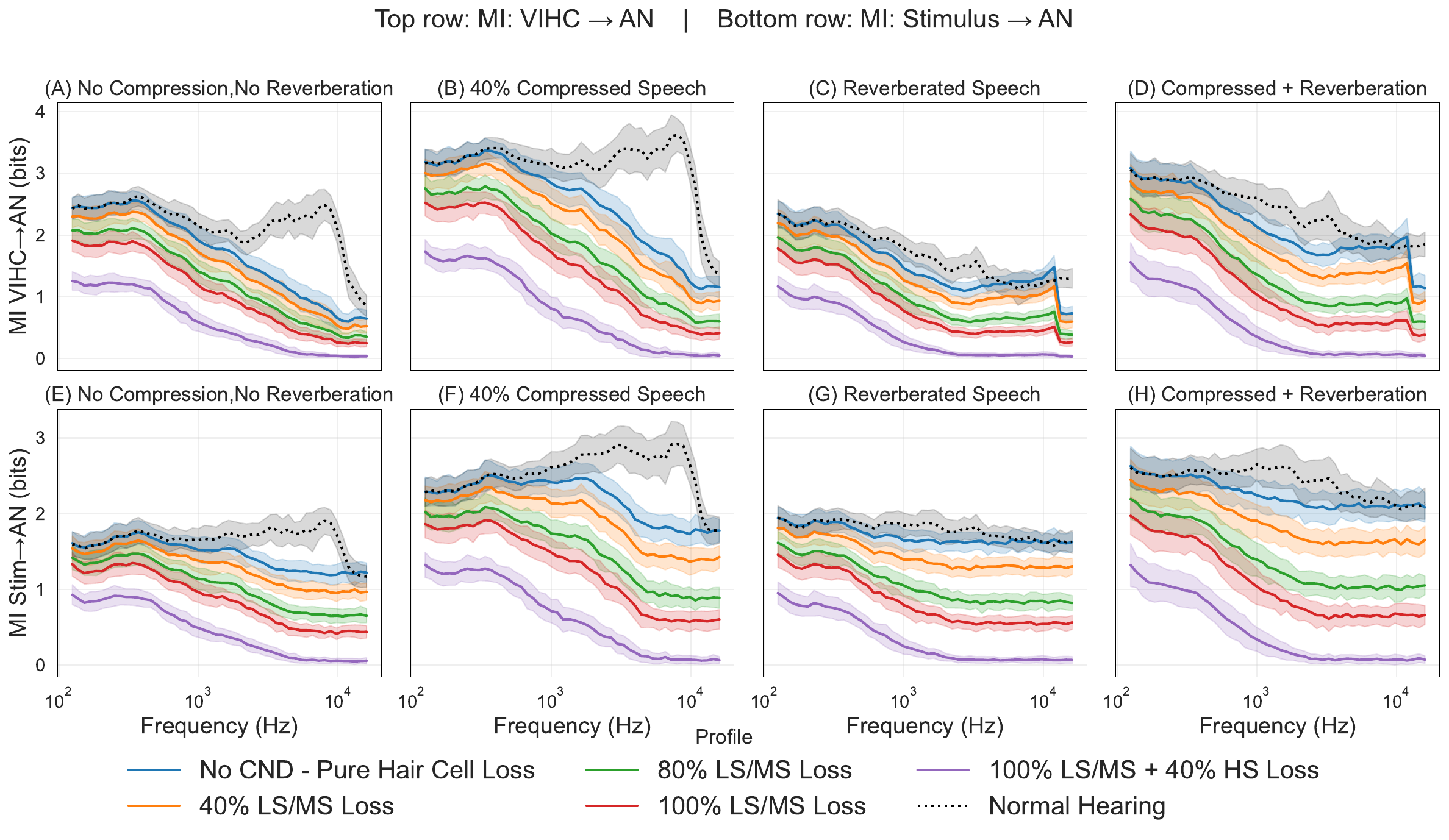} 
    \caption{Mutual information (MI) (Eq. \ref{eq:MI}) different hearing profiles under different speech material presented at 90 dB SPL.  Top row (A–D): MI between inner hair cell receptor potentials (VIHC) and auditory nerve fiber responses (ANF).  Bottom row (E–H): MI between acoustic stimulus and ANF responses. Probe conditions include: (A, E) clean speech (no compression, no reverberation), (B, F) 40\% time-compressed speech, (C, G) reverberated speech, and (D, H) combined compression + reverberation.}
    \label{fig:MI_curves_HL}
\end{figure}

\begin{figure}[t]
    \centering
    \includegraphics[width=1\linewidth]{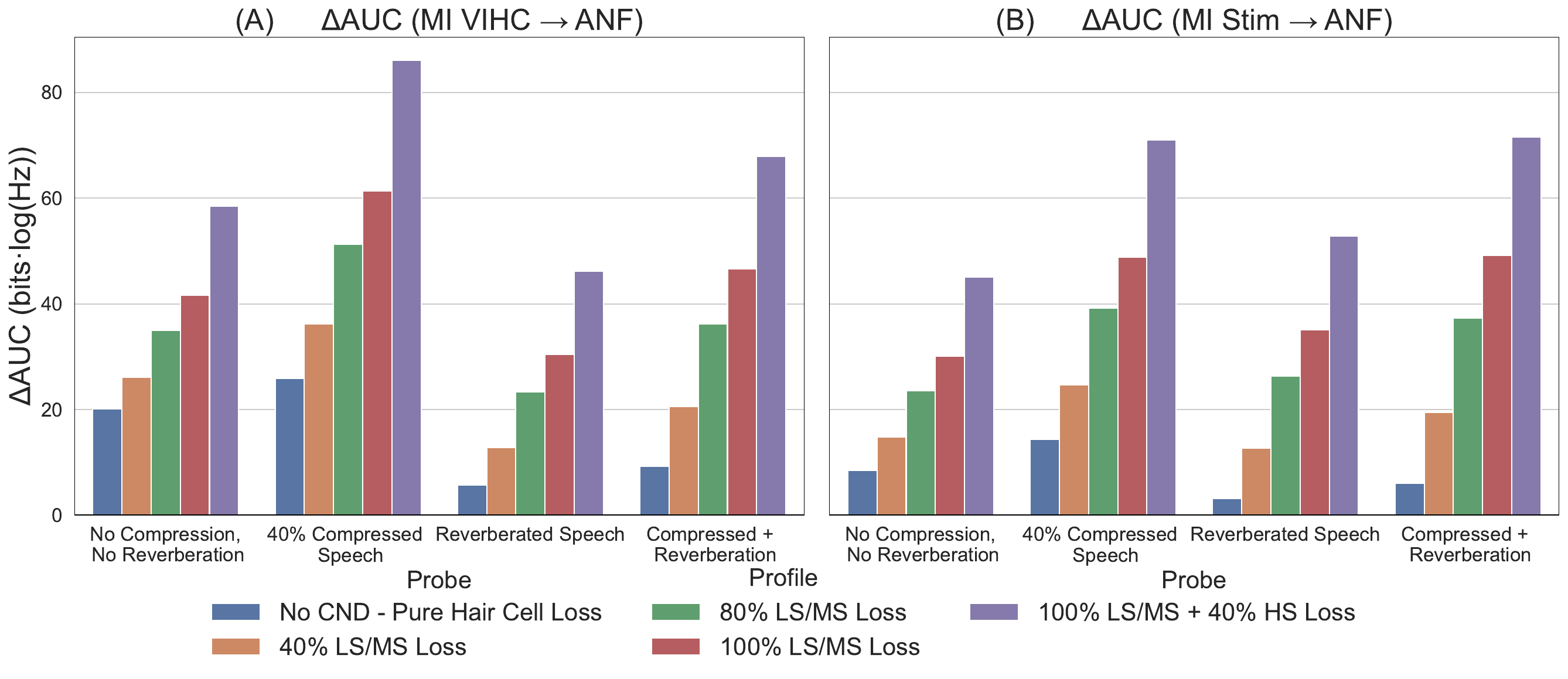} 
    \caption{Change in Area Under the Curve ($\Delta$AUC) (Eq. \ref{eq:deltaAUC}) for different profiles as listed in Table \ref{tab:CND_profiles} and probe conditions at 90 dB SPL. (A) $\Delta$AUC for MI calculated between inner hair cell receptor potentials (VIHC) and auditory nerve fiber responses (ANF). (B) $\Delta$AUC for MI calcualted between acoustic stimulus (Stim) and ANF responses.}
    \label{fig:normal_level}
\end{figure}

Figure \ref{fig:MI_curves_HL} shows the mean and STD of MI as defined by Eq. \ref{eq:Ivec} between $V_{IHC}$ and AN (top row) and between $X$ and AN (bottom row) calculated across the different corpus of speech (as defined in sec: \ref{subsec:speech_corpus}) at 90 dB SPL. We also calculated MI at other SPLs, but due to limitation of space only 90 dB SPL results are shown here. It can be seen that for normal hearing and no CND profile (dotted black) line, most amount of information is encoded across the high frequencies for both normal speech Fig. \ref{fig:MI_curves_HL}(A, E) and compressed speech Fig. \ref{fig:MI_curves_HL}(B, F), and blue line which represents the hearing loss profile as described in Table \ref{tab:audiometric_profile_A} and no CND, this information is lost at these high frequencies. As CND is added to this audiometric profile, the overall information begins to decrease across all type of speech materials. For the Reverberated speech Fig. \ref{fig:MI_curves_HL}(C, G), it can be seen that even for the normal hearing profile (dotted black) the overall information at high frequencies is lower that that observed in normal speech Fig. \ref{fig:MI_curves_HL}(A, E) and compressed speech Fig. \ref{fig:MI_curves_HL}(B, F). This same trend can also be seen for  compressed + reverberated speech Fig. \ref{fig:MI_curves_HL}(D, H) where the overall information at high frequencies is lower than the corresponding compressed speech profiles Fig. \ref{fig:MI_curves_HL}(B, F). 
Figure \ref{fig:normal_level} shows the overall information loss relative to normal hearing profile as defined by Eq. \ref{eq:deltaAUC} plotted for both $\mathbf{I}^{(V_{\mathrm{IHC}} \to \mathrm{AN})}$ (A) and $\mathbf{I}^{{(X} \to \mathrm{AN})}$ (B). It can be seen that across both conditions $\mathbf{I}^{(V_{\mathrm{IHC}} \to \mathrm{AN})}$ and  $\mathbf{I}^{{(X} \to \mathrm{AN})}$, as the CND increases the overall loss of information also increases, and for this loss is maximized across all profiles for the case of 40\% compressed speech. Reverberated speech material either produces lower loss than normal no compression no reverberation speech material or comparable level of loss across all profiles in both panels A and B.

\vspace{-10pt}
\section{Discussion}
\label{sec:discussion}
In this study, we introduce an information-theoretic framework to characterize the "difficulty" and sensitivity of different speech materials for detecting CND. By comparing mutual information (MI) between inner hair cell (IHC) receptor potentials and auditory nerve fiber (ANF) neurograms across normal-hearing and CND profiles, we sought to identify stimuli that maximize information loss in the presence of CND. The underlying assumption was that greater MI loss predicts poorer behavioral performance for individuals with CND, relative to normal-hearing listeners. Across all profiles analyzed, 40\% time-compressed speech produced the largest MI loss, confirming that temporally demanding material is especially sensitive to synaptic deficits. Introducing reverberation to compressed speech did not further increase the maximum information loss beyond that observed for compression alone. This effect can be explained by the fact that reverberation acts as a low-pass filter, smearing temporal fine structure and attenuating high-frequency consonant cues. Because the /CVC/ word lists used here and commonly used in both research and clinical testing rely heavily on these high-frequency cues, reverberation makes the task difficult for both normal-hearing and CND profiles, thereby reducing diagnostic specificity. Reverberation is ecologically relevant, challenging and demanding stimulus but since its degradations are not selective to synaptopathy, therefore using it as a probe can increase the likelihood of false positives in normal-hearing listeners with no CND. This is also in consistence with previous work \cite{cheema2025using} where compressed speech stimuli was found to be most sensitive using 2 dimensional correlation metric i.e. Neurogram Similarity Index Measure (NSIM).
Together, these results highlight that rapid, temporally dense speech stimuli—such as time-compressed words—offer the most sensitive and specific probes for revealing hidden hearing loss. Reverberation, while ecologically realistic, primarily reflects degradations that occur at the acoustic input stage and therefore should not serve as a primary diagnostic manipulation. By leveraging both Stim→AN and VIHC→AN information metrics, this framework provides a quantitative, mechanistic basis for designing speech probes that can reveal CND with high specificity.
\vspace{-10pt}
\section{Conclusion and Future Directions}
\label{sec:conclusion}
Using an information theoretic method, we were able to quantify the sensitivity of various types of speech material to detect CND. Our approach relied solely on the information encoding between inner hair cell receptors and the auditory nerve fibers, which may have a potential to assist the clinicians and scientists in evaluating the utility of speech tests designed to detect CND and associated perceptual deficits. In this work we looked at a single audiometric loss profile using a computational model. Future work could investigate information loss on speech task performance for subjects with various hearing loss profiles. Additionally, we only examined the response of auditory nerve to calculate information loss, future studies might evaluate responses from Inferior Colliculus (IC) and the mutual information degradation resulting from CND. The current models of auditory periphery can simulate the responses at IC \cite{Drakopoulos2024-md} and information theoretic analysis at the level of IC may reveal additional information about types of speech used for CND testing. Furthermore, future studies might utilize this approach to optimize parameters of hearing aids that can mitigate CND related information deficits.



\ninept
\bibliographystyle{IEEEbib}
\bibliography{strings,refs}

\end{document}